\documentclass[conference]{IEEEtran}
\ifCLASSINFOpdf
\else
\fi
\usepackage{graphicx}
\usepackage{multirow}
\usepackage[table]{xcolor}
\usepackage{amsmath} 
\usepackage{listings}
\usepackage{tikz}

\usetikzlibrary{arrows.meta, positioning}

\begin{document}
%
\title{Reducing Fragmentation and Starvation in GPU Clusters through Dynamic Multi-Objective Scheduling}

\author{
  \IEEEauthorblockN{Akhmadillo Mamirov}
  \IEEEauthorblockA{
    Department of Computer Science \\
    The College of Wooster\\
    amamirov26@wooster.edu
  }
}


%


\maketitle

\begin{abstract}
GPU clusters have become essential for training and deploying modern AI systems, yet real deployments continue to report average utilization near 50\%. This inefficiency is largely caused by fragmentation, heterogeneous workloads, and the limitations of static scheduling policies. This work presents a systematic evaluation of these issues and introduces three specialized dynamic schedulers: Hybrid Priority (HPS), Predictive Backfill (PBS), and Smart Batch (SBS). These schedulers are designed to improve utilization, fairness, and overall throughput in multi-tenant GPU clusters.

We evaluate all schedulers using a controlled simulation of 1,000 AI jobs on a 64-GPU, 8-node cluster that includes a realistic mix of training, inference, and research workloads. Static baselines (FIFO, SJF, Shortest, Shortest-GPU) achieve 45--67\% GPU utilization and 12.5--18.3 jobs per hour and experience severe starvation, with as many as 156 jobs waiting longer than 30 minutes. The dynamic schedulers significantly outperform these policies. HPS achieves the highest utilization (78.2\%), highest throughput (25.8 jobs per hour), and the lowest fairness variance among dynamic methods (457), reducing starvation to 12 jobs. PBS improves fragmentation handling and reaches 76.1\% utilization, while SBS increases efficiency for structurally similar jobs and reaches 74.6\% utilization.

Across all key metrics, including throughput, job wait times, fairness variance, and starvation, dynamic multi-objective schedulers consistently outperform single-objective heuristics. These results show that targeted and transparent scheduling strategies can meaningfully increase GPU efficiency in heterogeneous AI clusters and provide a practical foundation for future production scheduling frameworks.
\end{abstract}



%
\IEEEpeerreviewmaketitle

\section{Introduction}

The rapid growth of Artificial Intelligence (AI) and deep learning has elevated Graphics Processing Units (GPUs) to the center of modern computational infrastructure. Their massive parallelism enables efficient execution of training and inference workloads across domains such as autonomous driving, genomics, scientific computing, and large language models (LLMs). As organizations scale these workloads, investment in GPU clusters has surged. Yet despite this expansion, effectively utilizing these costly accelerators remains a persistent challenge.

In shared multi-tenant environments, job characteristics vary widely in duration, interactivity, and GPU demand, and traditional scheduling strategies often fail to adapt to this heterogeneity. As a result, GPUs frequently sit idle while jobs wait in queue, and resource fragmentation prevents large jobs from being scheduled even when sufficient aggregate capacity exists. Measurements from production cloud and enterprise clusters consistently report average utilization slightly above 50\%, far below what their hardware costs would seem to justify \cite{jeon}. This gap underscores the need for more adaptive, workload-aware scheduling techniques that can extract higher throughput, reduce queuing delays, and minimize fragmentation.

This research investigates these inefficiencies and develops new scheduling strategies aimed at improving GPU utilization in multi-tenant clusters. By examining how schedulers make placement decisions, how fragmentation emerges, and how workloads interact with heterogeneous hardware, this work seeks to establish practical and transparent methods for improving overall system performance.

\section{Background and Related Work}

\subsection{AI Workloads and Scheduling Policies}

Efficient scheduling must account for the diverse characteristics of AI workloads, which vary significantly in GPU demand, memory usage, and responsiveness requirements. Build sessions typically require instant and uninterrupted GPU access but exhibit low compute intensity. In contrast, model-training workloads are highly compute-intensive, often span multiple GPUs, and prioritize minimizing time-to-train. Inference workloads generally use fewer resources, with lower utilization and memory footprints compared to training jobs.

To overcome the limitations of static allocation, modern scheduling approaches incorporate fairness strategies and dynamic allocation techniques. Fairness-aware policies, such as Dominant Resource Fairness (DRF), attempt to allocate resources proportionally across competing workloads \cite{drf_fair_scheduling}. Dynamic schedulers adapt in real time based on workload behavior, cluster utilization, and global constraints.

A unique challenge arises from distributed training jobs that require gang scheduling, meaning all assigned GPUs must become available simultaneously and remain dedicated for the duration of the job. This requirement can cause significant queuing delays for large multi-GPU workloads and increases the risk of starvation under static policies \cite{wise_resource_sharing}.

\subsection{Fragmentation and Optimization Techniques}

A major challenge in multi-GPU clusters is fragmentation, which results in unused or underutilized resources even when the cluster appears heavily loaded.

\textbf{Node Fragmentation:}  
Occurs when available GPUs are distributed across multiple nodes, preventing large multi-GPU jobs from being scheduled.

\textbf{GPU Fragmentation:}  
Occurs when individual GPUs are only partially utilized because tasks consume limited fractions of memory or compute, leaving idle capacity.

Two primary strategies are commonly used to address fragmentation:

\begin{itemize}
    \item \textbf{Bin Packing:}  
    Reduces GPU-level fragmentation by placing smaller tasks onto partially utilized GPUs, increasing overall utilization.

    \item \textbf{Consolidation:}  
    Reduces node-level fragmentation by migrating smaller tasks to a subset of nodes, creating contiguous free GPU blocks on others for large jobs.
\end{itemize}

Several deep learning–oriented schedulers incorporate these ideas. Gandiva uses time-slicing and job swapping to improve responsiveness for interactive workloads \cite{gandiva}. Tiresias prioritizes short jobs by estimating remaining execution time\cite{tiresias}. Optimus builds online performance models to dynamically resize GPU allocations to minimize training time \cite{optimus}.  

These workload differences are further shaped by how modern deep learning models are trained across multiple GPUs \cite{dalmia}. Distributed training introduces communication overhead, synchronization requirements, and gang-scheduling constraints that directly influence cluster behavior and scheduling difficulty \cite{marble}. We summarize the major distributed training strategies below.

\subsection{Distributed Training in Modern Deep Learning}

Modern deep learning models frequently exceed the memory or compute capacity of a single GPU, requiring distributed training to scale effectively. Three primary forms of parallelism are widely used:

\textbf{Data Parallelism (DP):}  
Each GPU holds a full model replica and processes a different mini-batch of data. Gradients are aggregated across GPUs using an All-Reduce operation, after which parameters are synchronized. DP is simple and scales well but introduces communication overhead at large batch sizes \cite{anyscale2025}.

\textbf{Model Parallelism (MP):}  
Used when a model cannot fit within one GPU's memory. The model is partitioned across devices, and each GPU computes a portion of the forward and backward pass. Variants include layer-wise, operator-wise, and tensor parallelism. MP reduces memory pressure but requires frequent communication of intermediate activations \cite{anyscale2025}.

\textbf{Pipeline Parallelism:}  
Extends MP by dividing the model into sequential stages and feeding micro-batches through a pipeline. This improves utilization by overlapping computation across stages, though synchronization at the end of each iteration introduces pipeline flush overhead \cite{pipeline_huang}.

\textbf{Training Optimization Techniques:}  
Several methods improve performance and efficiency in distributed training:
\begin{itemize}
    \item \textit{Batch size scaling} for higher hardware efficiency \cite{crossbow}.
    \item \textit{Mixed precision training} to reduce memory use and accelerate compute \cite{nvidia_mixed_precision}.
    \item \textit{Gradient accumulation} to simulate large batches on memory-limited hardware \cite{gradient_accumulation}.
    \item \textit{ZeRO} to partition optimizer and model states across GPUs \cite{zeroMemory}.
    \item \textit{CPU offloading} to reduce GPU memory pressure \cite{cpu_offloading}.
\end{itemize}

Our work builds on these concepts by integrating multiple strategies such as efficiency-driven selection, aging-based fairness, and predictive backfilling into a set of specialized, transparent, and modular scheduling algorithms tailored for diverse AI workloads.

\section{Baseline Schedulers and Motivation for New Designs}

\subsection{Design Goals Summary}

 Based on observed failure modes and system requirements, we identified four core goals for effective scheduling:

\begin{itemize}
    \item \textbf{Fairness:} Prevent starvation and ensure bounded wait times across heterogeneous job types.
    \item \textbf{Fragmentation Reduction:} Improve placement so GPUs are used contiguously rather than scattered.
    \item \textbf{Throughput Improvement:} Increase completed jobs per hour by minimizing idle GPU time.
    \item \textbf{Predictable Behavior:} Produce stable, explainable scheduling decisions under changing workloads.
\end{itemize}

These goals guided both our analysis of existing methods and the design of the specialized schedulers introduced later.

\subsection{Baseline Schedulers}

Before developing new scheduling strategies, we implemented and evaluated four baseline schedulers that represent common single-objective heuristics used in GPU clusters:

\begin{itemize}
    \item \textbf{FIFO} schedules jobs strictly by arrival order. It guarantees fairness in ordering but suffers from low utilization when long jobs block many short ones.
    \item \textbf{SJF} prioritizes jobs requiring the fewest GPUs. This improves throughput for small tasks but causes systematic starvation of large training jobs.
    \item \textbf{Shortest (SRTF)} selects the job with the smallest remaining time. It increases responsiveness but repeatedly postpones long-running workloads.
    \item \textbf{Shortest-GPU} selects the job with minimum remaining GPU-time (duration \(\times\) GPU count). This balances time and resource use but still disadvantages large jobs when many small tasks arrive.
\end{itemize}

\subsection{Limitations of Static Policies}

Evaluation of these schedulers revealed three consistent structural limitations:

\begin{itemize}
    \item \textbf{Starvation:} SJF and SRTF repeatedly delayed large or long jobs whenever smaller tasks were available.
    \item \textbf{Fragmentation:} FIFO and SJF often left GPUs idle because they did not consider resource fit or cluster topology.
    \item \textbf{Lack of Adaptability:} None of the schedulers adjusted priorities based on queue length, workload composition, or system load.
\end{itemize}

These limitations show that real GPU workloads require schedulers that consider \emph{multiple interacting factors}—time, resource demand, fairness, and fragmentation—rather than optimizing a single metric. This motivates the development of the specialized multi-objective schedulers described in the next section.

\subsection{Failure of the Adaptive Multi-Factor Scheduler}

To address the limitations of static heuristics, we designed an adaptive multi-factor scheduler intended to
balance three objectives: \emph{efficiency} (favor short or low-cost jobs), \emph{fairness} (prevent starvation through
aging), and \emph{resource awareness} (account for GPU demand and fragmentation). Each job was assigned scores
for these three components, and dynamically adjusted weights were applied based on system conditions such
as queue length.

However, empirical evaluation showed that this approach was unstable and difficult to tune:

\begin{itemize}
    \item \textbf{Objective Interference}: Small changes in weights caused drastic shifts in scheduling behavior,
    making the system unpredictable.
    \item \textbf{Normalization Sensitivity}: Score normalization produced counterintuitive effects, sometimes
    causing long jobs to receive higher efficiency scores than short ones.
    \item \textbf{Binary Threshold Effects}: Dynamic reweighting created abrupt changes in behavior when the
    queue crossed preset thresholds.
    \item \textbf{Maintenance Burden}: The scoring model became difficult to debug, reason about, or trust in a
    production-like environment.
\end{itemize}

In short, attempting to solve all objectives simultaneously through a unified scoring function introduced
more complexity than benefit. The scheduler exhibited unstable prioritization patterns and was highly
sensitive to parameter tuning, confirming that overly general solutions can be brittle under heterogeneous
GPU workloads.

\subsection{Transition to Specialized Multi-Objective Schedulers}

The failure of the adaptive multi-factor scheduler informed the next phase of the design process. Instead of
building a universal scoring model, we developed \emph{specialized} schedulers, each targeting a well-defined failure
mode observed in the baselines:

\begin{itemize}
    \item \textbf{Hybrid Priority Scheduler (HPS)} addresses fairness and starvation while incorporating
    efficiency and GPU blocking awareness.
    \item \textbf{Predictive Backfill Scheduler (PBS)} targets fragmentation by selecting jobs that fit resource
    gaps and prioritizing efficient job placements.
    \item \textbf{Smart Batch Scheduler (SBS)} exploits job similarity to reduce overhead, reuse execution
    contexts, and improve throughput.
\end{itemize}

Each scheduler pursues a narrow objective rather than attempting to solve all scheduling
criteria simultaneously. This specialization produced policies that were simpler, explainable, tunable, and
more robust in evaluation. The remainder of this section presents the design of these schedulers and the
observed improvements over the baseline methods.

\section{Methodology}

\begin{figure}[t]
    \centering
    \begin{tikzpicture}[
        node distance=0.9cm,
        rounded corners,
        every node/.style={
            draw, 
            minimum width=3.2cm,
            minimum height=0.85cm,
            font=\footnotesize
        },
        arrow/.style={-Latex, thick}
    ]

    \node (workload) {Workload};
    \node (scheduler) [below=of workload] {Scheduler};
    \node (decision) [below=of scheduler] {Placement Decision};
    \node (metrics) [below=of decision] {Metrics};

    \draw[arrow] (workload) -- (scheduler);
    \draw[arrow] (scheduler) -- (decision);
    \draw[arrow] (decision) -- (metrics);

    \end{tikzpicture}

    \caption{Evaluation pipeline: workloads are submitted to the scheduler, which makes placement decisions that generate measurable performance metrics.}
    \label{fig:pipeline-overview}
\end{figure}
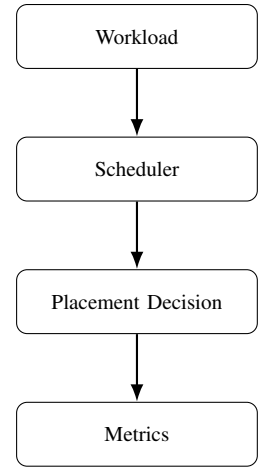

\subsection{Simulation Environment}

All schedulers were evaluated using identical job streams, cluster
configurations, and random seeds to ensure fair comparison.

The overall evaluation flow is illustrated in Fig.~\ref{fig:pipeline-overview}, which shows how workloads pass through each scheduler to generate placement decisions and performance metrics.

\textbf{Cluster Configuration:}  
The simulated environment consisted of 8 nodes with 8 NVIDIA GPUs per node, totaling 64 GPUs.  
Cluster state was reset before each scheduler run to ensure consistent initial conditions.

\textbf{Workload Generation:}  
Each experiment submitted 1{,}000 jobs following a statistically defined mixed workload representative of
multi-tenant AI clusters. Job mixes were generated using fixed random seeds and validated to match the
intended distribution.

\textbf{Job Types:}
\begin{itemize}
    \item Inference: 50\%
    \item Training: 30\%
    \item Research / Experimental: 20\%
\end{itemize}

\textbf{GPU Requirements:}
\begin{itemize}
    \item 1 GPU: 35\%
    \item 2 GPUs: 25\%
    \item 4 GPUs: 20\%
    \item 8 GPUs: 15\%
    \item 16+ GPUs: 5\%
\end{itemize}

\textbf{Job Durations:}
\begin{itemize}
    \item Short: 40\% (less than 30 minutes)
    \item Medium: 35\% (30 minutes to 2 hours)
    \item Long: 20\% (2--8 hours)
    \item Very long: 5\% (greater than 8 hours)
\end{itemize}

\subsection{Schedulers Evaluated}

We benchmarked seven schedulers in total:

\textbf{Static Schedulers:}
\begin{itemize}
    \item FIFO (arrival-order)
    \item SJF (min GPU count)
    \item Shortest (min remaining time)
    \item Shortest-GPU (min GPU-time product)
\end{itemize}

These single-objective heuristics provide predictable behavior but lack adaptability under heterogeneous workloads.

\textbf{Dynamic Schedulers:}
\begin{itemize}
    \item Hybrid-Priority (HPS): combines efficiency scoring with aging to prevent starvation.
    \item Predictive-Backfill (PBS): uses forward-looking gap analysis and resource forecasting.
    \item Smart-Batch (SBS): groups compatible jobs to improve packing and shared execution efficiency.
\end{itemize}

Table~\ref{tab:schedulers} summarizes each scheduler and its core objective.

\subsection{Evaluation Metrics}

Scheduler performance was assessed across multiple metric categories:

\textbf{Performance Metrics:}  
Throughput (jobs/hour), average wait time, job completion time (JCT), GPU utilization, and resource efficiency.

\textbf{Fairness Metrics:}  
Variance in wait times, starvation count (jobs waiting $>$ 30 minutes), minimum/maximum wait times, and success rate.

\textbf{System-Level Metrics:}  
Makespan, resource fragmentation, queue length evolution, and preemption or conflict events.

\subsection{Testing Procedure}

Evaluation followed a four-phase process:

\textbf{1. Workload Generation.}  
Job sets were generated using statistical models with fixed seeds and validated for distributional accuracy.

\textbf{2. Scheduler Execution.}  
Each scheduler ran independently on the same workload. Cluster state was reinitialized for every run.  
Multiple trials were conducted to reduce variance and confirm stability.

\textbf{3. Data Collection.}  
Real-time metrics (GPU utilization, queue length), job-level statistics (wait time, execution time), and  
system-level indicators (throughput, fragmentation, starvation events) were logged throughout execution.

\textbf{4. Analysis.}  
Collected data were aggregated using means and variances, and confidence intervals were computed.  
Direct scheduler comparisons were performed, and correlations between metrics were identified. . 

\begin{table}[t]
\centering
\setlength{\tabcolsep}{3pt}
\begin{tabular}{l l l}
\hline
\textbf{Scheduler} & \textbf{Heuristic} & \textbf{Objective} \\
\hline

FIFO & Arrival order & Fairness / simplicity \\

SJF & Shortest time & Throughput \\

Shortest & Min GPU count & Concurrency \\

Shortest-GPU & Min (GPU $\times$ time) & Balanced efficiency \\

\rowcolor{green!20} HPS & Multi-factor + aging & Efficiency + fairness \\

\rowcolor{green!20} PBS & Efficiency + backfill & Utilization, low fragmentation \\

\rowcolor{green!20} SBS & Similarity grouping & Throughput, synergy \\

\hline
\end{tabular}
\vspace{0.5em}
\caption{Schedulers and their core design principles.}
\label{tab:schedulers}
\end{table}

\section{Specialized Scheduler Implementation}

The dynamic schedulers (HPS, PBS, and SBS) were designed to address the specific failure modes of static policies, particularly starvation and fragmentation.

\subsection{Hybrid Priority Scheduler (HPS)}

The Hybrid Priority Scheduler (HPS) was developed after observing that simple heuristics (e.g., FIFO, SJF)
produce predictable starvation patterns, while multi-factor adaptive schedulers become difficult to tune and
unstable under heterogeneous workloads. HPS adopts a middle-ground approach: it combines three well-understood
mechanisms---efficiency bias, aging-based fairness, and GPU-blocking mitigation---into a compact and transparent scoring model.

\subsubsection*{Scoring Model}

HPS assigns each job a composite multiplicative score:
\[
\text{Score} = \text{BaseScore} \times \text{AgingScore} \times \text{GPUPenalty}.
\]

\textbf{BaseScore (Efficiency):}  
Prioritizes shorter jobs using a smoothed inverse-time function:  
\[
\text{BaseScore} = \frac{1}{1 + \text{remaining\_time}/3600}.
\]

\textbf{AgingScore (Fairness):}  
Boosts jobs that exceed the aging threshold. If
$\text{wait\_time} > \text{aging\_threshold}$,  
\[
\text{AgingScore} =
\text{aging\_boost} \cdot 
\min\!\left(\text{wait\_time}/\text{max\_wait\_time},\, 1\right);
\]
otherwise $\text{AgingScore} = 1$.

\textbf{GPUPenalty (Blocking Mitigation):}  
Penalizes large GPU allocations to reduce blocking:
\[
\text{GPUPenalty} = \frac{1}{1 + \text{num\_gpu}/4}.
\]

This multiplicative structure guarantees that no single factor dominates, and each reflects a targeted design:  
efficiency for small jobs, aging for fairness, and GPU-penalization for throughput and fragmentation control.

\subsubsection*{Implementation}

Default parameters were chosen to provide predictable behavior across mixed workloads:
\begin{itemize}
    \item \texttt{aging\_threshold = 300s} (begin aging after 5 minutes)
    \item \texttt{aging\_boost = 2.0} (upper bound on fairness amplification)
    \item \texttt{max\_wait\_time = 1800s} (cap for aging influence)
\end{itemize}

\subsubsection*{Performance Characteristics}

HPS consistently improves throughput compared to FIFO and improves fairness relative to SJF and Shortest.
Moderate GPU penalization reduces blocking of small jobs while ensuring that large multi-GPU jobs eventually
advance through aging. This balance improves cluster utilization without requiring complex model tuning.

\subsubsection*{Limitations and Trade-Offs}

HPS requires parameter tuning (\texttt{aging\_threshold}, \texttt{boost}, \texttt{max\_wait\_time}) to match workload
distribution, and remains more complex than single-objective heuristics. However, the simplicity of its scoring
rule avoids the instability and interpretability issues of earlier multi-factor schedulers while achieving
measurable gains in fairness, utilization, and starvation prevention.

\subsection{Predictive Backfill Scheduler (PBS)}

The Predictive Backfill Scheduler (PBS) uses forward-looking heuristics to increase GPU utilization by
selecting jobs that maximize efficiency while reducing fragmentation. Unlike reactive schedulers, PBS
evaluates estimated runtimes, GPU requirements, and potential resource gaps to make decisions that
maintain consistent throughput.

\subsubsection*{Selection Principles}

PBS applies three decision rules and a fallback:

\begin{enumerate}
    \item \textbf{Efficiency Priority.}
    Jobs are ranked by efficiency, defined as work per GPU per unit time.
    The most efficient job is selected only if it is at least $(1+\tau)$ more efficient than the next best
    candidate, with $\tau = 0.1$ as the stability threshold. This prevents minor efficiency fluctuations
    from dominating scheduling choices.

    \item \textbf{Gap Filling.}
    If no job exceeds the efficiency threshold, PBS selects a ``small'' job whose GPU requirement is
    below a minimum threshold~$\gamma$. Among these, the job with the shortest remaining time is chosen
    to fill fragmented GPU slots and increase concurrency.

    \item \textbf{Blocking Avoidance.}
    When no small job is appropriate, PBS selects a medium-duration job (remaining time below~$T$)
    with the smallest GPU footprint. This reduces cases where long, high-GPU jobs monopolize the
    cluster.

    \item \textbf{Fallback.}
    If none of the above apply, PBS selects the job with the shortest remaining runtime to ensure
    deterministic behavior.
\end{enumerate}

\subsubsection*{Rationale and Behavior}

The efficiency rule allocates resources to jobs that yield the highest productivity, but only when the
difference is meaningful. The threshold $\tau = 0.1$ filters out noise: differences smaller than 10\% do not
affect scheduling. Gap-filling prevents idle GPUs by selecting jobs that fit existing resource fragments, while
the blocking rule improves long-term throughput by avoiding resource monopolization.

\subsubsection*{Predictive Backfill Extension}

PBS also evaluates whether pairs of jobs can run concurrently. A pair $(j_1, j_2)$ is considered feasible if:

\begin{itemize}
    \item Their combined GPU demand does not exceed node capacity.
    \item Their runtimes are compatible within a tolerance~$\Delta$, preventing one job from completing
    too early and leaving GPUs idle.
\end{itemize}

For feasible pairs, PBS computes a combined efficiency score based on total work over total GPU-time and
selects the pair with the highest score. This predictive backfilling enables higher utilization than
single-job heuristics, especially in mixed workloads with diverse GPU requirements.

\subsection{Smart Batch Scheduler (SBS)}

The Smart Batch Scheduler (SBS) improves throughput by grouping compatible jobs into batches that can
share kernels, memory layouts, and preprocessing steps. This reduces context switching and increases GPU
utilization in workloads.

\subsubsection*{Batch Formation}

Jobs are first grouped by model family or structural similarity. A batch $B$ is considered feasible if it satisfies:
\[
\sum_{j \in B} \text{num\_gpu}(j) \le G_{\max},
\]
where $G_{\max}$ is the GPU capacity allocated for batching, and
\[
\text{Sim}(B) \ge \theta,
\]
where $\theta$ is the similarity threshold.  
The similarity score uses variance in duration and GPU usage:
\[
\text{Sim}(B) = \frac{1}{1 + \sigma_t^2 + \sigma_g^2},
\]
with $\sigma_t^2$ and $\sigma_g^2$ denoting variance of remaining times and GPU counts.

\subsubsection*{Batch Scoring}

Each feasible batch is scored using combined efficiency:
\[
\text{Eff}(B) = \frac{\sum_{j \in B} \text{iterations}(j)}
                     {\left(\sum_{j \in B} \text{num\_gpu}(j)\right)\cdot 
                      \max_{j \in B} \text{remaining\_time}(j)}.
\]

The final batch score multiplies efficiency by similarity:
\[
\text{Score}(B) = \text{Eff}(B)\cdot \text{Sim}(B).
\]

The batch with the highest score provides the selected job.

\subsubsection*{Fallback Strategy}

If no batch satisfies feasibility or scoring requirements, SBS selects an individual job using a reduced form of
the same scoring criteria, emphasizing efficiency and low GPU demand. This ensures stable behavior even
when batching opportunities are scarce.

\subsubsection*{Rationale and Behavior}

SBS increases utilization by:
\begin{itemize}
    \item reducing context switching for similar jobs,
    \item enabling reuse of kernels and memory structures,
    \item aligning job durations to reduce idle GPU periods,
    \item preventing fragmentation through coordinated scheduling.
\end{itemize}

When workloads contain repeated architectures or parameter sweeps, SBS consistently forms high-scoring
batches that improve throughput. When workloads are heterogeneous, the fallback mode performs similarly
to single-job schedulers, avoiding regressions.

Relative to HPS and PBS, SBS has higher computational overhead due to batch discovery, but it provides
unique efficiency gains through synergy exploitation that cannot be achieved with purely individual-job
heuristics.

\section{Evaluation and Analysis}

\subsection{Static Scheduler Limitations}

Static policies optimized for a single metric consistently failed to balance performance and fairness under mixed workload conditions.

\begin{itemize}
    \item \textbf{FIFO:} Achieved the highest fairness score (126 wait time variance) but had the lowest throughput (12.5 jobs/hour) and GPU utilization (45.2\%), leading to significant resource wastage as long jobs blocked the queue.

    \item \textbf{SJF:} Improved utilization (67.4\%) but suffered from extreme starvation, recording 156 starved jobs and the lowest fairness score (2847 variance) by systematically delaying large workloads.

    \item \textbf{Shortest-GPU:} Achieved better efficiency by balancing time and resource consumption, but still recorded 67 starved jobs, confirming that static heuristics cannot fully mitigate starvation risks in heterogeneous queues.
\end{itemize}

\subsection{Starvation Analysis}

Starvation was evaluated by counting all jobs that waited more than 
30 minutes before receiving GPU resources. 

Static schedulers exhibited severe starvation under mixed workloads. 
SJF delayed large jobs aggressively, resulting in 156 starved tasks, 
and Shortest and Shortest-GPU also recorded high starvation counts 
(89 and 67, respectively). These behaviors confirm that single-metric 
policies systematically deprioritize certain job classes whenever 
short or low-GPU jobs are abundant.

In contrast, dynamic schedulers substantially reduced starvation. 
Hybrid-Priority limited starvation to 12 jobs by applying aging 
after prolonged wait times. Predictive-Backfill kept starvation low 
(18 jobs) by filling gaps with small tasks without permanently 
delaying large ones. Smart-Batch also reduced starvation compared to 
static policies, though batching constraints produced moderately 
higher starvation (25 jobs) than HPS.

Success-rate measurements follow the same pattern: static schedulers 
complete 79–87\% of jobs, while dynamic schedulers consistently exceed 
94\% completion. These results demonstrate that starvation in static 
policies is structural, whereas fairness mechanisms in dynamic 
schedulers provide predictable and bounded waiting behavior.

\section*{Performance Metrics: Throughput, Wait Time, Utilization, and Fairness}

Beyond starvation outcomes, we compared schedulers across
four key metrics: throughput, average wait time, GPU utilization, and fairness variance.
These results reinforce the limitations of static schedulers and highlight the multi-objective
benefits achieved by the dynamic approaches.

FIFO provides the lowest wait-time variance (126) because it strictly honors arrival order,
but it also delivers the lowest throughput (12.5 jobs/hour) and GPU utilization (45.2\%),
demonstrating that fairness alone cannot compensate for inefficient resource usage. SJF
achieves relatively high utilization (67.4\%) but exhibits extreme unfairness with a variance
of 2847, since short jobs repeatedly displace long training workloads. Shortest and Shortest-GPU
improve throughput relative to FIFO, yet their fairness variances (1957 and 1678) confirm
that optimizing a single dimension---time or GPU footprint---still leads to imbalanced queue
progression.

In contrast, the dynamic schedulers achieve consistently better multi-objective performance as can be seen from Table \ref{tab:scheduler-performance}.
Hybrid-Priority (HPS) attains the highest throughput (25.8 jobs/hour), highest utilization
(78.2\%), and the lowest fairness variance among dynamic methods (457), validating the
effectiveness of combining efficiency scoring with controlled aging. Predictive-Backfill (PBS)
achieves competitive throughput while reducing fragmentation, and Smart-Batch (SBS)
further improves fairness (variance 679) by grouping compatible jobs, allowing them to
progress through the queue in a more synchronized manner.

Fairness variance is computed using the standard statistical definition:
\[
\text{Variance} = 
\frac{\sum_{i=1}^{n} (x_i - \mu)^2}{n},
\]
where $x_i$ denotes the wait time of job $i$ and $\mu$ is the mean wait time. Lower
variance indicates a more equitable distribution of waiting times across all jobs, which is
essential in multi-tenant GPU clusters.

\begin{table}[th]
\centering
\setlength{\tabcolsep}{4pt}
\begin{tabular}{l c c c c c}
\hline
\textbf{Scheduler} &
\textbf{Jobs/hr} &
\textbf{GPU Util.} &
\textbf{Wait (s)} &
\textbf{Fairness} &
\textbf{Starved} \\
\hline
Hybrid Priority (HPS)     & 25.8 & 78.2\% & 757 & 457 & 12 \\
Predictive Backfill (PBS) & 24.3 & 76.1\% & 823 & 524 & 18 \\
Smart Batch (SBS)         & 23.7 & 74.6\% & 891 & 679 & 25 \\
\hline
\end{tabular}
\vspace{0.5em}
\caption{Performance comparison of HPS, PBS, and SBS under mixed workloads.}
\label{tab:scheduler-performance}
\end{table}

\section{Conclusion}
This research systematically investigated why GPU utilization in multi-tenant AI clusters remains far below hardware capacity and demonstrated how scheduling decisions directly shape throughput, fairness, and fragmentation. By evaluating 1,000 mixed AI workloads on a 64-GPU cluster, we confirmed that static, single-objective schedulers—FIFO, SJF, Shortest, and Shortest-GPU—struggle to balance competing goals. They achieved only 45--67\% utilization, produced high fairness variance (up to 2847), and recorded severe starvation, with as many as 156 jobs waiting more than 30 minutes. These results highlight the structural limitations of policies that optimize only for arrival order, job size, or remaining time.

The three specialized dynamic schedulers introduced in this study demonstrate how focused design choices can overcome the shortcomings of static heuristics. Hybrid Priority (HPS) achieved the strongest overall performance, delivering 78.2\% utilization, 25.8 jobs/hour throughput, and a fairness variance of 457 while reducing starvation to 12 jobs. Predictive Backfill (PBS) improved resource placement by filling fragmented gaps and maintaining stable throughput at 24.3 jobs/hour. Smart Batch (SBS) leveraged similarity-based grouping to reduce overhead and increase concurrency, particularly in workloads with repeated architectures. Together, these schedulers reduced starvation by an order of magnitude compared to static heuristics and produced far more balanced waiting-time distributions.

A broader conclusion emerging from this study is that effective GPU scheduling does not require large, monolithic optimizers. Instead, the most reliable and interpretable improvements come from targeted, modular mechanisms that address specific failure modes: aging for fairness, predictive gap filling for fragmentation, and similarity grouping for synergy exploitation. This modular perspective provides a practical and extensible framework for future cluster schedulers.

Looking ahead, the scheduling strategies introduced here form a foundation for integrating hardware-aware placement, adaptive parallelism, preemptive checkpointing, and multi-cluster coordination. As AI workloads continue to grow in scale and heterogeneity, transparent and specialized multi-objective scheduling will remain essential for maximizing the value of GPU infrastructure.



%
\nocite{*}
\bibliography{main}

@inproceedings{drf_fair_scheduling,
author = {Ghodsi, Ali and Zaharia, Matei and Hindman, Benjamin and Konwinski, Andy and Shenker, Scott and Stoica, Ion},
title = {Dominant resource fairness: fair allocation of multiple resource types},
year = {2011},
publisher = {USENIX Association},
address = {USA},
booktitle = {Proceedings of the 8th USENIX Conference on Networked Systems Design and Implementation},
pages = {323–336},
numpages = {14},
location = {Boston, MA},
series = {NSDI'11}
}

@inproceedings{jeon,
author = {Jeon, Myeongjae and Venkataraman, Shivaram and Phanishayee, Amar and Qian, unjie and Xiao, Wencong and Yang, Fan},
title = {Analysis of large-scale multi-tenant GPU clusters for DNN training workloads},
year = {2019},
isbn = {9781939133038},
publisher = {USENIX Association},
address = {USA},
booktitle = {Proceedings of the 2019 USENIX Conference on Usenix Annual Technical Conference},
pages = {947–960},
numpages = {14},
location = {Renton, WA, USA},
series = {USENIX ATC '19}
}

@misc{wise_resource_sharing,
      title={Scheduling Deep Learning Jobs in Multi-Tenant GPU Clusters via Wise Resource Sharing}, 
      author={Yizhou Luo and Qiang Wang and Shaohuai Shi and Jiaxin Lai and Shuhan Qi and Jiajia Zhang and Xuan Wang},
      year={2024},
      eprint={2407.13088},
      archivePrefix={arXiv},
      primaryClass={cs.DC},
      url={https://arxiv.org/abs/2407.13088}, 
}

@inproceedings{optimus,
author = {Peng, Yanghua and Bao, Yixin and Chen, Yangrui and Wu, Chuan and Guo, Chuanxiong},
title = {Optimus: an efficient dynamic resource scheduler for deep learning clusters},
year = {2018},
isbn = {9781450355841},
publisher = {Association for Computing Machinery},
address = {New York, NY, USA},
url = {https://doi.org/10.1145/3190508.3190517},
doi = {10.1145/3190508.3190517},
booktitle = {Proceedings of the Thirteenth EuroSys Conference},
articleno = {3},
numpages = {14},
location = {Porto, Portugal},
series = {EuroSys '18}
}

@inproceedings{tiresias,
author = {Gu, Juncheng and Chowdhury, Mosharaf and Shin, Kang G. and Zhu, Yibo and Jeon, Myeongjae and Qian, Junjie and Liu, Hongqiang and Guo, Chuanxiong},
title = {Tiresias: a GPU cluster manager for distributed deep learning},
year = {2019},
isbn = {9781931971492},
publisher = {USENIX Association},
address = {USA},
booktitle = {Proceedings of the 16th USENIX Conference on Networked Systems Design and Implementation},
pages = {485–500},
numpages = {16},
location = {Boston, MA, USA},
series = {NSDI'19}
}

@inproceedings{gandiva,
author = {Chaudhary, Shubham and Ramjee, Ramachandran and Sivathanu, Muthian and Kwatra, Nipun and Viswanatha, Srinidhi},
title = {Balancing efficiency and fairness in heterogeneous GPU clusters for deep learning},
year = {2020},
isbn = {9781450368827},
publisher = {Association for Computing Machinery},
address = {New York, NY, USA},
url = {https://doi.org/10.1145/3342195.3387555},
doi = {10.1145/3342195.3387555},
booktitle = {Proceedings of the Fifteenth European Conference on Computer Systems},
articleno = {1},
numpages = {16},
location = {Heraklion, Greece},
series = {EuroSys '20}
}

@article{pipeline_huang,
  author       = {Yanping Huang and
                  Yonglong Cheng and
                  Dehao Chen and
                  HyoukJoong Lee and
                  Jiquan Ngiam and
                  Quoc V. Le and
                  Zhifeng Chen},
  title        = {GPipe: Efficient Training of Giant Neural Networks using Pipeline
                  Parallelism},
  journal      = {CoRR},
  volume       = {abs/1811.06965},
  year         = {2018},
  url          = {http://arxiv.org/abs/1811.06965},
  eprinttype    = {arXiv},
  eprint       = {1811.06965},
  bibsource    = {dblp computer science bibliography, https://dblp.org}
}

@online{anyscale2025,
  author    = {Anyscale},
  title     = {What is Distributed Training?},
  year      = {2025},
  url       = {https://www.anyscale.com/blog/what-is-distributed-training},
  note      = {Accessed: 2025-02-07},
}

@misc{zeroMemory,
      title={ZeRO: Memory Optimizations Toward Training Trillion Parameter Models}, 
      author={Samyam Rajbhandari and Jeff Rasley and Olatunji Ruwase and Yuxiong He},
      year={2020},
      eprint={1910.02054},
      archivePrefix={arXiv},
      primaryClass={cs.LG},
      url={https://arxiv.org/abs/1910.02054}, 
}

@inproceedings{cpu_offloading,
author = {Xu, Dong and Feng, Yuan and Shin, Kwangsik and Kim, Daewoo and Jeon, Hyeran and Li, Dong},
title = {Efficient Tensor Offloading for Large Deep-Learning Model Training based on Compute Express Link},
year = {2024},
isbn = {9798350352917},
url = {https://doi.org/10.1109/SC41406.2024.00100},
doi = {10.1109/SC41406.2024.00100},
booktitle = {Proceedings of the International Conference for High Performance Computing, Networking, Storage, and Analysis},
articleno = {94},
numpages = {18},
location = {Atlanta, GA, USA},
series = {SC '24}
}

@INPROCEEDINGS{gradient_accumulation,
  author={Huang, Zimeng and Jiang, Bo and Guo, Tian and Liu, Yunzhuo},
  booktitle={2023 IEEE/ACM 23rd International Symposium on Cluster, Cloud and Internet Computing (CCGrid)}, 
  title={Measuring the Impact of Gradient Accumulation on Cloud-based Distributed Training}, 
  year={2023},
  volume={},
  number={},
  pages={344-354},
  doi={10.1109/CCGrid57682.2023.00040}}

@misc{nvidia_mixed_precision,
  author       = {NVIDIA},
  title        = {Mixed Precision Training},
  year         = {2025}, 
  url          = {https://docs.nvidia.com/deeplearning/performance/mixed-precision-training/index.html},
  note         = {Accessed: 2025-02-07}  
}

@article{crossbow,
author = {Koliousis, Alexandros and Watcharapichat, Pijika and Weidlich, Matthias and Mai, Luo and Costa, Paolo and Pietzuch, Peter},
title = {Crossbow: scaling deep learning with small batch sizes on multi-GPU servers},
year = {2019},
issue_date = {July 2019},
publisher = {VLDB Endowment},
volume = {12},
number = {11},
issn = {2150-8097},
url = {https://doi.org/10.14778/3342263.3342276},
doi = {10.14778/3342263.3342276},
journal = {Proc. VLDB Endow.},
month = jul,
pages = {1399–1412},
numpages = {14}
}

@article{dalmia,
  author={Dalmia, Preyesh and Mahapatra, Rohan and Intan, Jeremy and Negrut, Dan and Sinclair, Matthew D.},
  journal={IEEE Transactions on Parallel and Distributed Systems}, 
  title={Improving the Scalability of GPU Synchronization Primitives}, 
  year={2023},
  volume={34},
  number={1},
  pages={275-290},
  doi={10.1109/TPDS.2022.3218508}
}

@INPROCEEDINGS{marble,
  author={Han, Jingoo and Rafique, M. Mustafa and Xu, Luna and Butt, Ali R. and Lim, Seung-Hwan and Vazhkudai, Sudharshan S.},
  booktitle={2020 20th IEEE/ACM International Symposium on Cluster, Cloud and Internet Computing (CCGRID)}, 
  title={MARBLE: A Multi-GPU Aware Job Scheduler for Deep Learning on HPC Systems}, 
  year={2020},
  volume={},
  number={},
  pages={272-281},
  doi={10.1109/CCGrid49817.2020.00-66}}
\bibliographystyle{IEEEtran}

\end{document}